\documentstyle[times,epsfig]{jaa}

\def \kms    {\mbox{$\rm kms^{-1} $ }}
\def \cm3    {\mbox{$\rm cm^{-3}$}}
\def \ne     {\mbox{$ n_{e}~$}}

\def \Te     {$ T_e $ }
\def \Tr     {$ T_{R100} $ }
\def \H0     {{$\rm H^{\circ} $} }
\def \HI    {\mbox{\rm H{\hspace{0.5mm}\scriptsize I }}}
\def \HII    {\mbox {\rm H{\hspace{0.5mm}\scriptsize II }} }
\def \CII    {\mbox {\rm C{\hspace{0.5mm}\scriptsize II } }}
\def \delV   {\mbox{$\rm \Delta V $} }
\def \Vlsr   {\mbox{$\rm V_{lsr} $ }}
\def \bn     {\mbox{$ b_n $} }

\def \Itau   {\mbox {$\rm \frac {\int T_ld\nu}{T_{sys}} $} }

\def \SB79   {\mbox {$ b_n \rightarrow 1 $ as $ \rm n \rightarrow \infty $ }}
\def \GN89  {\mbox{$ b_n \rightarrow 0 $ as $ \rm n \rightarrow n_{critical} $ }}
\def \ie     {{\it i.e.~}}
\def \eg     {{\it e.g.~}}
\def \et     {{\it et al.}}

\def \beqn   {\begin{equation} }
\def \eeqn   {\end{equation} }

\def \mJyb     {{$\rm mJy~beam^{-1}$}}

\begin{document}

\title[Carbon Recombination Lines]
{Carbon Recombination Lines from the Galactic Plane at 34.5 \& 328 MHz } 

\author[Kantharia \& Anantharamaiah]{N. G. Kantharia,$^1$
   \thanks{e-mail: ngk@ncra.tifr.res.in} K. R. Anantharamaiah, $^2$
   \thanks{e-mail: anantha@rri.res.in} \\
 $^1$National Centre for Radio Astrophysics (TIFR), 
 Pune 411007, India \\ 
 $^2$Raman Research Institute   
 Bangalore 560080, India}

\maketitle

\begin{abstract}
We present results of a search for carbon recombination lines in 
the Galaxy at 34.5 MHz (C$575\alpha$) made using the dipole array at
Gauribidanur near Bangalore.  Observations made towards 32 directions,
led to detections of lines in absorption at nine positions.  Followup
observations at 328 MHz (C$272\alpha$) using the Ooty Radio Telescope
detected these lines in emission.  A VLA D-array observation of one
of the positions at 330 MHz yielded no detection implying a lower limit
of 10' for the angular size of the line forming region.

The longitude-velocity distribution of the observed carbon lines indicate
that the line forming region are located mainly between 4 kpc and 7 kpc from
the Galactic centre.  Combining our results with published carbon recombination
line data near
76 MHz (\nocite{erickson:95} Erickson \et  1995) we obtain constraints
on the physical parameters of the line forming regions.  We find that if the
angular size of the line forming regions is $\ge 4^{\circ}$, then the range
of parameters that fit the data are: \Te $= 20-40$ K, \ne $\sim 0.1-0.3$ \cm3
and pathlengths $\sim 0.07-0.9$ pc which may correspond to thin photo-dissociated
regions around molecular clouds.  On the other hand, if the line forming
regions are $\sim 2^{\circ}$ in extent, then warmer gas (\Te $\sim 60-300$~K)
with lower electron densities (\ne $\sim 0.03-0.05$ \cm3 ) extending over several
tens of parsecs along the line of sight and possibly associated with
atomic \HI gas can fit the data.  Based on the range of
derived parameters, we suggest that the carbon line regions are most likely 
associated with photo-dissociation regions
\end{abstract}

\begin{keywords}
interstellar medium: clouds, lines, line profiles, radio lines
\end{keywords}

\section{Introduction}

Radio recombination lines have been extensively used to investigate the ionized component
of the interstellar medium.   Recombination lines of hydrogen, helium and
carbon have been
unambiguously identified in the spectra obtained towards \HII regions.  
Since the ionization potentials
of hydrogen (13.6 eV) and helium (24.4 eV) are relatively high, these elements are 
ionized by the strong UV radiation in the vicinity of O and B stars.  
Therefore radio recombination lines of H \& He generally
trace hot ionized regions ($T_e \sim$ 5000 - 10000 K).  On the other hand, 
the first and second ionization potential of the fourth 
most abundant element, carbon are 11.4 eV and $\sim 24$ eV.  
Carbon is likely to be doubly-ionized inside the \HII regions. 
However, lower energy photons (E $<$ 13.6 eV) which escape from the \HII regions can 
singly-ionize carbon both in the immediate vicinity of the \HII region 
(if the \HII region is ionization
bounded) as well as in the interstellar clouds which may be farther away. 
In these regions where hydrogen, helium and oxygen are largely neutral and where carbon and
few other elements with lower ionization potential (e.g. sulphur) are singly ionized, the
temperatures are likely to be much lower i.e. less than a few hundred K.
Radio recombination lines of singly-ionized carbon, thus,  
generally trace cooler regions of the interstellar medium. 

The \CII regions detectable in recombination lines can be categorized into two types.
The first type are the ionized-carbon regions which occur in the immediate vicinity of 
some bright \HII regions such as W3 and NGC~2024.  We shall refer to such regions as
classical \CII regions since these are carbon Stromgren spheres around a central star.
Carbon recombination lines were first detected from 
such classical \CII regions by 
\nocite{palmer:67}Palmer \et (1967) at a frequency of $\sim 5$ GHz and have
subsequently been studied at $\nu > 1$ GHz towards many \HII regions
using both single dishes and interferometers 
(e.g. \nocite{pankonin:77} Pankonin \et 1977, \nocite{van:80} van Gorkom \et 1980, 
\nocite{roelfsema:87} Roelfsema \et 1987, 
\nocite{onello:95} Onello \& Phillips 1995, 
\nocite{kantharia:98a} Kantharia \et 1998a).  Carbon recombination  lines from 
classical \CII regions are narrow ($ 4-10 $ \kms) and are sometimes
comparable in strength to the hydrogen recombination line observed from 
the associated hot \HII region. 
Both the narrow width and intensity of the line 
suggest that the carbon line originates in a cooler ($<$ few hundred K) region where
stimulated emission of the background continuum is likely to cause enhanced line emission. 
Classical \CII regions are generally observed in carbon recombination lines
at frequencies $\nu \ge 1$ GHz. To our knowledge no carbon recombination line
from classical \CII regions has been detected below 1 GHz.

A second class of \CII regions, which we will refer
to as diffuse \CII regions, and which this paper is mainly concerned with, 
were discovered by \nocite{konovalenko:80} Konovalenko \& Sodin (1980).  
Using the UTR-2 low-frequency
radio telescope in Ukraine, \nocite{konovalenko:80}Konovalenko and Sodin (1980) observed
an absorption line at 26.13 MHz in the direction of the strong radio source Cas A.
This low-frequency absorption feature was later correctly identified by 
\nocite{blake:80}Blake, 
Crutcher and Watson (1980) as the C$631\alpha$ recombination line arising
in a region at a temperature of $\sim 50$ K and electron density 
$\sim 0.1$ \cm3 .  The region was tentatively identified with the 
diffuse neutral \HI clouds in the Perseus arm.  Since then,
the direction of Cas A has been observed in recombination lines of carbon at frequencies
ranging from 14 MHz to 1420 MHz (\nocite{payne:89} Payne \et  1989,  
\nocite{payne:94} Payne \et 1994 and the references therein).
The most recent results (\nocite{kantharia:98b}Kantharia \et 1998b)
suggest that the diffuse \CII region in the Perseus arm towards Cas~A 
could be at a temperature of $\sim$75 K with an 
electron density of $\sim$0.02 \cm3 , which supports the identification of
these regions with neutral \HI clouds.  Interferometric observations of the spatial
distribution of the C272$\alpha$ carbon line (near 332 MHz) over the face of Cas~A also
supports this association (\nocite{anantha:94} Anantharamaiah \et 1994, 
\nocite{kantharia:98b} Kantharia \et 1998b).  The
low-frequency recombination lines from diffuse \CII region towards Cas~A exhibit
interesting characteristics.  The lines which are in absorption below 150 MHz turnover
into emission above 200 MHz.  Moreover, the width of the lines increase dramatically towards
lower frequencies because of pressure and radiation broadening.  The variation
of line strength and line width with frequency are important diagnostics of
the physical conditions in the line forming regions. 
  
If the diffuse \CII regions which are detected in low-frequency recombination
lines of carbon towards Cas~A are associated with the neutral \HI component of the 
interstellar medium, then
it may be expected that such low-frequency lines are a widespread phenomenon.
Detections have indeed been made in several other directions.  Carbon recombination lines  
were detected in absorption near 25 MHz from the directions of G75.0+0.0  
\& NGC 2024 by \nocite{konovalenko:84a} Konovalenko (1984a) and towards L1407, DR21
\& S140 by \nocite{golynkin:91} Golynkin \& Konovalenko (1991).  \nocite{anantha:88}
Anantharamaiah \et (1988) detected absorption lines
near 75 MHz towards M16 and the Galactic Center.  With hindsight, it now appears that the
carbon lines detected in emission near 327 MHz using the Ooty Radio Telescope 
towards 14 directions in the galactic plane by \nocite{anantha:85} Anantharamaiah (1985) belong to this 
category.  The first major fruitful search for low-frequency
recombination lines of carbon  was conducted by \nocite{erickson:95}Erickson \et (1995)
 at 76 MHz using the Parkes radio telescope in Australia.
Absorption lines of carbon were detected from all the observed positions with longitudes 
$<20^{\circ}$ in the Galactic plane.  These observations showed that diffuse \CII regions 
are a common phenomenon in the inner Galaxy.  

In this paper, we present observations of recombination lines of carbon  around 34.5 MHz 
made using the low-frequency dipole array at Gauribidanur (which is near Bangalore, India)  
and around 328 MHz made using the Ooty Radio Telescope (ORT). 
Out of the 32 directions, most of them in the galactic plane, that 
were searched at 34.5 MHz,  lines were
detected in absorption towards nine directions.  
At 327 MHz, twelve positions were observed and
lines were detected in emission towards seven of these.  Six of the detected lines are
common to 34.5 MHz and 328 MHz.  To obtain an estimate of the size of the line-forming region, 
observations were made with two angular resolutions ($\sim 2^{\circ} \times 2^{\circ}$ 
and $\sim 2^{\circ} \times 6'$) using the  ORT. 
We also followed up one of the detections with observations near 332 MHz using the 
D-Configuration of the Very Large Array.  All the six directions toward which lines were 
detected here at 34.5 MHz and 328 MHz, have also been detected at 76 MHz by 
\nocite{erickson:95} Erickson \et (1995). 
In this paper, we combine the results at these three frequencies to constrain the nature of the line 
forming region.
 
This paper is organized as follows. In Section 2, the observations at 34.5 MHz using
the Gauribidanur dipole array, at 327 MHz using the Ooty Radio Telescope and at 332
MHz using the Very Large Array are described and the results are presented.  In Section 3,
we  obtain constraints on the sizes of the line forming regions using line ratios observed with
different angular resolutions.  The probable location of the line forming region in the
galactic disk  is constrained  using an $l-v$ diagram of the observed lines.  In Section 3, 
we also derive constraints on the combination of electron density, temperature and radiation
field surrounding the line forming region using the variation of line width with frequency.
Section 4 is devoted to modelling the line forming region by fitting the observed variation of
integrated optical depth with frequency.  In Section 5, the results of modelling are
discussed in the light of what we know about the interstellar medium.  Finally the paper is
summarized in Section 6.

\section{Observations and Results} 

\subsection {Gauribidanur Observations near 34.5 MHz}

The low-frequency dipole array at Gauribidanur which operates near 34.5
MHz (\nocite{deshpande:89}Deshpande \et 1989 and references therein) was used during 
July$-$October 1995 to observe several $\alpha-$transitions($\Delta$n = 1) with principal 
quantum numbers ranging from n = 570 to 580.  The Gauribidanur radio telescope is a
meridian transit instrument consisting of 1000 broadband dipoles arranged
in a T$-$shaped configuration.  The East-West (EW) arm consists of 640
dipoles, distributed over a distance of 1.38 km whereas the North-South
arm extends southwards from the centre of the EW arm and consists of 360
dipoles laid over a distance of 0.45 km.  The present set of observations
were carried out using the EW arm in total power mode, which gave an angular
resolution of $\sim 21'\times25^\circ$ ($\alpha \times \delta$) at zenith. The
effective collecting area of the EW arm is 160$\lambda^2$ ($\lambda$=8.67 m).

The Gauribidanur telescope has limited tracking facility 
(\nocite{deshpande:89}Deshpande \et, 1989) 
which enables the array to track a source with declination $\delta$, for a period of
$40$sec$\delta$ minutes about its transit time.  
Since these low-frequency recombination lines are weak, each position would
have required several weeks of observations with the Gauribidanur telescope to reach the
desired sensitivity.
In order to reduce the total duration of the observations, we employed a
multi-line receiver in which eight different transitions with $\Delta n=1$ near $n=575$  
were observed simultaneously and the spectra averaged.  
The details of the multi-line receiver and the observing procedure have been
described in \nocite{kantharia:98b} Kantharia \et (1998b).

The data collected were carefully examined and all those affected by
interference were removed.  The final spectrum was obtained by averaging all
the observed transitions.  The spectra are hanning smoothed and therefore have a 
resolution of 0.5 kHz.  The total bandwidth is 32 kHz ($\sim 250$ \kms).

\begin{table}
\begin{center}
\caption{\sf List of Sources that were searched for Carbon Radio
Recombination Lines at 34.5 MHz}
\vspace{0.3cm}
\begin{tabular}{lccccc}
   & & & & & \\
\hline
{\bf Position} & $\bf \alpha(1950)$ & $\bf \delta(1950)$ & {\bf Central Vel}
& $\bf \frac{\Delta T_{rms}}{T_{sys}}$ & {\bf Effective}  \\
  & & & {\bf Setting} & & {\bf Int Time}\\
  & hh mm ss & $^{\circ}~~'~~''$ & \kms & $\times 10^{-3}$ & hours \\ 
\hline
 G342+00 & 16 50 34 & -43 37 38 & -20.0 & 0.35  & 63.5 \\
 G352+00 & 17 21 47 & -35 37 25  & -20.0 & 0.20 & 95.3 \\
 G00+00  & 17 42 27  & -28 55 00 & 0.0 & 0.20 & 58.7 \\  
 G05+00  & 17 54 00 & -24 37 59 & 0.0 & 0.23 & 70.3 \\
 G10+00  & 18 04 47 & -20 17 51 & 0.0 & 0.20 & 91.3 \\
 G14+00  & 18 12 59  & -16 48 00 & 20.0  & 0.22 & 55.0 \\
 G16.5+00  & 18 17 57 & -14 36 18 & 25.0 & 0.22 & 68.3 \\
 G25+00   & 18 34 12 & -7 05 51 & 45.0 & 0.23 &  78.3   \\
 G45.2+00  & 19 11 44 & 10 48 50 & 20.0 & 0.23 & 45.8   \\
 G50+00  & 19 21 02  & 15 02 41 & 25.0 & 0.27 &  108.0   \\
 G55+00  & 19 31 08 & 19 25 32 & 25.0 & 0.23  & 50.7   \\
 G63+00  & 19 48 31 & 26 21 12  & 20.0 & 0.19 & 72.0 \\
 G75+00  & 20 19 02  & 36 26 46 & 0.0 & 0.15 & 188.0 \\
 G81+00  & 20 37 17  & 41 16 58 & 25.0 & 0.31  & 114.0   \\
 G85+00  & 20 23 08 & 21 22 05 & 25.0 & 0.34 &  41.7  \\
 G97+00  & 21 42 21 & 52 56 10 & -20.0 & 0.29 &  91.5  \\
 G99+00  & 21 52 47 & 54 11 59 & -20.0 & 0.33 &  169.5  \\
 G100+00  & 21 58 15 & 54 48 34 & -20.0 & 0.20 & 178.3   \\
 G125+00  & 01 06 22 & 62 31 58 & -20.0 & 0.37 & 88.8   \\
 G130+00  & 01 48 45 & 61 47 14 & -20.0 & 0.37 & 128.5   \\
 G145+00  & 03 34 08 & 55 24 12 & -20.0 & 0.37 & 100.0   \\
 S 140     & 22 17 36 & 63 04 00 & -20.0 & 0.39 & 24.3   \\
 DR 21     & 20 37 13  & 42 09 00 & 0.0  & 0.35 & 28.3 \\
 Orion    & 05 32 48 & -05 27 00 & 0.0  & 0.54 & 19.7   \\
 Cygnus Loop  & 20 49 30 & 29 50 00 & 25.0 & 0.16 & 99.8    \\
 Cygnus A  & 19 57 45 & 40 36 00 & 0.0 & 0.23 & 62.3   \\
 W3       & 02 21 50 & 61 53 20 & -50.0 & 0.39  & 238.8   \\
 W49      & 19 08 51 & 09 02 27 & 60.0 & 0.46 & 42.9   \\
 W51      & 19 20 17 & 14 02 01 & 50.0 & 0.20 & 69.0   \\
 G203.1+2.1   & 06 38 17 & 09 43 20 & 8.0 &  0.5  & 12.0   \\
 G224.6-2.4   & 07 01 56 & -11 23 55 & 15.0 & 0.22 & 49.1   \\
 Rosette Nebula & 06 29 18 & 04 57 00 & 0.0 & 0.34 & 37.5   \\
\hline
\end{tabular}

\label{tab1}
\end{center}
\end{table}

We observed 32 positions, most of these being in the Galactic plane. 
Table \ref{tab1} gives the observational parameters.
The observable positions were limited by the transit nature of the
Gauribidanur telescope.  Within the declination range 
accessible to Gauribidanur telescope ($-45^{\circ}$ to $75^{\circ}$),
the chosen positions were determined partly by the strong Galactic background in the 
inner Galaxy and partly by the
positive results of other low-frequency observations (\nocite{konovalenko:84a} 
Konovalenko 1984a, \nocite{anantha:85} Anantharamaiah 1985, 
\nocite{golynkin:91} Golynkin \& Konovalenko 1991).  Carbon recombination lines were detected in absorption
from nine of these directions.  At such high quantum levels, collisions thermalize 
the level populations
and since the brightness temperature of the non-thermal background
radiation field is much higher than the temperature of the thermal gas, 
the lines appear in absorption.  The spectra obtained towards nine of the 
directions are shown in Fig \ref{fig1}. 
The results from a high signal-to-noise ratio spectrum obtained towards 
Cas~A have been presented in a separate publication 
(\nocite{kantharia:98b}Kantharia \et 1998b).  
The peak line-to-continuum ratios observed towards the
nine directions shown in Fig {\ref{fig1} are of the order 
of a few times $10^{-4}$.  Gaussians were fitted to the detected 
spectra and the parameters are listed in Table \ref{tab2}.
No lines were detected 
from the other positions to a 5$\sigma$ limit of $5.0 \times 10^{-4}$.
In almost all the cases, only a linear baseline had to be 
removed to obtain the final spectra.

\begin{figure}
\epsfig{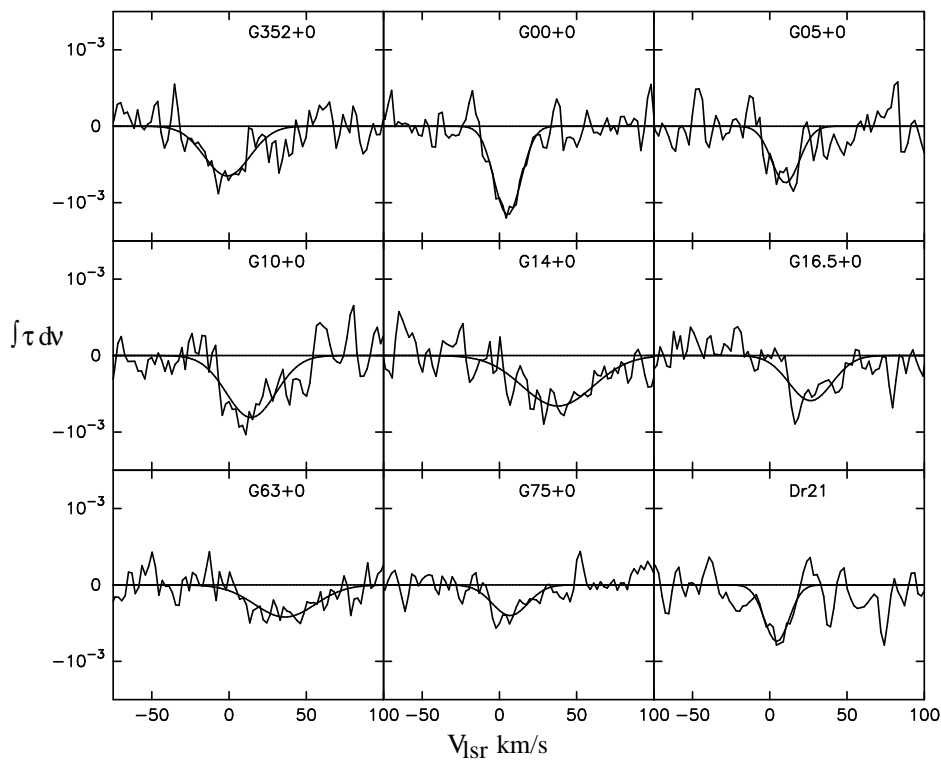}
\vspace{0.5cm}
\caption{\sf Carbon recombination lines detected near 34.5 MHz ($n\sim 575$).
The smooth curve superposed on the observed spectrum is
the Gaussian fit to the line profile.}
\label{fig1}
\end{figure}

A relatively strong absorption line of carbon was detected from the 
direction of the Galactic centre (G$00+00$) as shown in Fig \ref{fig1}.  
Carbon recombination lines have been observed from this direction
in absorption at 76 MHz (\nocite{anantha:88} Anantharamaiah \et 1988, 
\nocite{erickson:95} Erickson \et 1995),  
42 MHz (\nocite{smirnov:96} Smirnov \et 1996) and in
emission at 408 MHz (\nocite{pedlar:78} Pedlar \et 1978) and 328 MHz (\nocite{anantha:85} Anantharamaiah 1985, 
\nocite{roshi:97} Roshi \& Anantharamaiah, 1997).  

\begin{table}
\caption{\sf Parameters of the lines detected at 34.5 MHz.
 $\bf \Delta V $ is the FWHM.
The quoted uncertainties are $1\sigma$. }
\vspace{0.5cm}
\begin{tabular}{clccccc}
\hline
{\bf No.} &{\bf Source} & $T_{l}/T_{sys}$ & \Vlsr & $\Delta V $  &  \Itau & $T_{rms}/T_{sys}$ \\
 & & $\times 10^{-3}$ & \kms & \kms & $s^{-1}$ & $\times 10^{-3}$ \\
\hline 
 1 & G352+0 & $-0.65(0.07)$ & $-1.2(1.8)$ & $34.8(2.6)$ & $-3.3(0.3)$ & 0.21 \\ 
  & & & & & &\\ 
 2 & G00+00 & $-1.16(0.09)$ & $5.3(0.8)$ & $20.5(1.1)$ & $-3.5(0.3)$ & 0.21 \\ 
&  & & & & & \\ 
 3 & G05+00 & $-0.74(0.10)$ & $10.2(1.4)$ & $21.2(2.0)$ & $-2.3(0.3)$ & 0.26 \\ 
&  & & & & & \\ 
 4 & G10+00 & $-0.81(0.08)$ & $14.3(1.7)$ & $37.5(2.5)$ & $-4.3(0.2)$ & 0.29  \\ 
 4a & G10+00 & $-0.93(0.10)$ & $9.6(1.5)$ & $24.9(2.3)$ & $-4.7(0.4)$ & 0.28\\ 
   &        & $-0.47(0.11)$ & $40.2(2.6)$ & $18.4(3.7)$ &  &  \\
& & & & & & \\
 5 & G14+00 & $-0.66(0.07)$ & $37.8(2.7)$ & $54.0(3.8)$ & $-5.1(0.4)$ & 0.22 \\
& & & & & & \\ 
 6 & G16.5+0& $-0.59(0.07)$ & $26.4(2.0)$ & $32.8(2.8)$ & $-2.8(0.3)$ & 0.22 \\ 
& & & & & & \\ 
 7 & G63+00 & $-0.42(0.06)$ & $36.2(3.2)$ & $46.0(4.5)$ & $-2.8(0.4)$ & 0.21 \\ 
& & & & & & \\ 
 8 & G75+00 & $-0.40(0.05)$ & $6.9(1.8)$ & $27.1(2.6)$ & $-1.5(0.2)$ & 0.18\\ 
& & & & & & \\ 
 9 & DR 21   & $-0.74(0.13)$ & $4.5(1.6)$ & $18.8(2.2)$ & $-2.0(0.3)$ & 0.26 \\ 
\hline
\end{tabular}

\label{tab2}
\end{table}

Carbon lines were detected from all the six observed directions
in the inner galactic plane ($l= 352^{\circ} - 17^{\circ}$). 
This result is similar to that of \nocite{erickson:95} Erickson \et (1995) 
who detected carbon recombination 
lines near 76 MHz from all their observed positions in the 
longitude range $342^{\circ}$ to $20^{\circ}$.
We also detected lines near 34.5 MHz from DR 21, G63+00 and G75+00. 
The lines detected are either at positive velocities or close to 0 \kms.  
The line-to-continuum ratios of these lines range from 
a few times $10^{-4}$ to $10^{-3}$ and the
width-integrated line-to-continuum ratio ranges from 1.5 to 4 $\rm s^{-1}$.
The lines have widely varying widths with the narrowest lines 
($\sim 19$ \kms) occuring towards {\mbox DR 21} 
and the broadest lines ($\sim 54$ \kms) towards G14+00. 
\nocite{erickson:95} Erickson \et (1995) have also observed line widths 
(near 76 MHz) ranging from 5 \kms~ to 47 \kms. 

The carbon line observed towards G10+00 appears to be composed of two components. 
Parameters derived from a double
component Gaussian fit to this profile are listed
in Table \ref{tab2}.  A similar trend is also seen in the profiles towards G05+00 and G14+00.
However, the signal-to-noise is not adequate for a double component fit. 

The rms noise and the effective integration time on the spectra with no detection
are listed in Table \ref{tab1}.  The spectra were smoothed to a spectral 
resolution corresponding to a typical
line width ($\sim 20$ \kms) and the spectra were again examined for 
the presence of a spectral feature. 
A few of the spectra seem to suggest the presence of a weak 
signal $e.g.$ G342+00, G55+00 and G99+00.
However further observations are required to confirm these.  

\subsection {Ooty Observations Near 328 MHz}

The observations with the ORT (\nocite{swarup:71} Swarup \et, 1971) were carried out 
in two sessions: March-April 1995 and October 1995.  
The ORT is a $30$m$\times530$m (EW $\times$ NS) parabolic cylinder with an equatorial mount.
The length of the telescope along the NS is divided into 11 north and 11 south modules. 
Since the main aim of these observations was to search for the
emission counterpart of the carbon lines seen at 34.5 MHz, a subset of the 
positions observed at 34.5 MHz were observed with the ORT.  Since the 
size of the carbon line-forming regions is not known, we undertook 
observations in two modes of operation of the ORT which yielded two different
angular resolutions ($\sim 2^{\circ}\times6'$ and $\sim 2^{\circ}\times2^{\circ}$ 
at zenith).

In the first mode, 12 positions were observed with the resolution of the 
entire telescope ({\it i.e.} $ \sim 2^{\circ} \times 6'$) whereas in the
second mode, 10 of these positions were investigated with the resolution 
of a single module ({\it i.e.} $ \sim 2^{\circ} \times 2^{\circ}$).
The telescope was used in the total power mode and each position was observed for
at least 6 hours.  In the higher resolution mode, data are collected 
simultaneously from two beams (which are labelled beams 5 and 7),
separated by 6.6' in declination.  Four successive recombination line
transitions (C$270\alpha$ to C$273\alpha$) were observed simultaneously from both the beams
using a dual 4-line reciever developed for Galactic recombination line
observations (\nocite{roshi:2000}Roshi \& Anantharamaiah, 2000). 
The four transitions were averaged to get the final spectra, 
which on the average have an effective integration time $\sim30$ hours. 
A spectral resolution of 2.1 \kms was obtained for each band after hanning smoothing. 
All the lines were detected in emission. 

\begin{table}
\begin{center}
\caption{\sf Positions searched for Carbon recombination lines at 328 MHz using the ORT}
\vspace{0.3cm}
\begin{center}
\begin{tabular}{lcccccc}
\hline
& & & \multicolumn{2}{c}{\large \bf Full Telescope } & 
\multicolumn{2}{c}{\large \bf Single Module } \\ 
{\bf Source} & $\alpha(1950)$ & $ \delta(1950)$ 
    & $\bf \frac{T_{rms}}{T_{sys}}$ & $\bf t_{eff}$ 
    & $\bf \frac{T_{rms}}{T_{sys}}$ & $\bf t_{eff}$   \\
 & hh mm ss & $^{\circ}~~'~~''$ & $10^{-3}$ & hrs & $10^{-3}$ & hrs  \\ 
\hline
 G355+00 & 17 29 54 & -33 08 01  & 0.19 & 29.0 & 0.13 & 34.7 \\
 G00+00  & 17 42 27  & -28 55 00 & 0.16 & 23.0 & 0.11 & 34.5 \\  
 G05+00  & 17 54 00 & -24 37 59 & 0.15  & 20.5 & 0.13 & 34.4 \\
 G10+00  & 18 04 47 & -20 17 51 & 0.17 & 25.7 & 0.13 &  37.1 \\
 G14+00  & 18 12 59  & -16 48 00  & 0.20 & 19.3 & 0.19 & 20.1\\
 G16.5+00 & 18 17 57 & -14 36 18 &0.19 & 33.9 & 0.15 & 31.2 \\
 G30+00   & 18 43 29 & -02 39 48 & 0.26 & 16.0 & - & -  \\
 G50+00  & 19 21 02  & 15 02 41 & - & - & 0.17 & 28.2      \\
 G62+00  & 19 46 15 & 25 29 40  & 0.19 & 26.1 & 0.15 & 33.9\\
 W49     & 19 08 51 & 09 02 27 & 0.20 & 20.7  & 0.17 & 42.9  \\
 W51     & 19 20 17 & 14 02 01 & 0.20 & 14.5  & 0.15 & 26.5 \\
 G75+00  & 20 19 02 & 36 26 46 & 0.20 & 30.7 & - & -\\
 NGC2024 & 05 39 11 & -01 55 50 & 0.20 & 14.5 & - & -   \\  
\hline
\end{tabular}
\end{center}

\label{tab3}
\end{center}
\end{table}

Table \ref{tab3} shows the observational parameters.
On the average, the rms noise on the spectra
obtained using the full ORT was $\rm T_{rms}/T_{sys} \sim 2\times10^{-4}$ and on the
spectra obtained using a single module $\rm T_{rms}/T_{sys} \sim 1.5\times10^{-4}$.

\begin{figure}
\epsfig{file=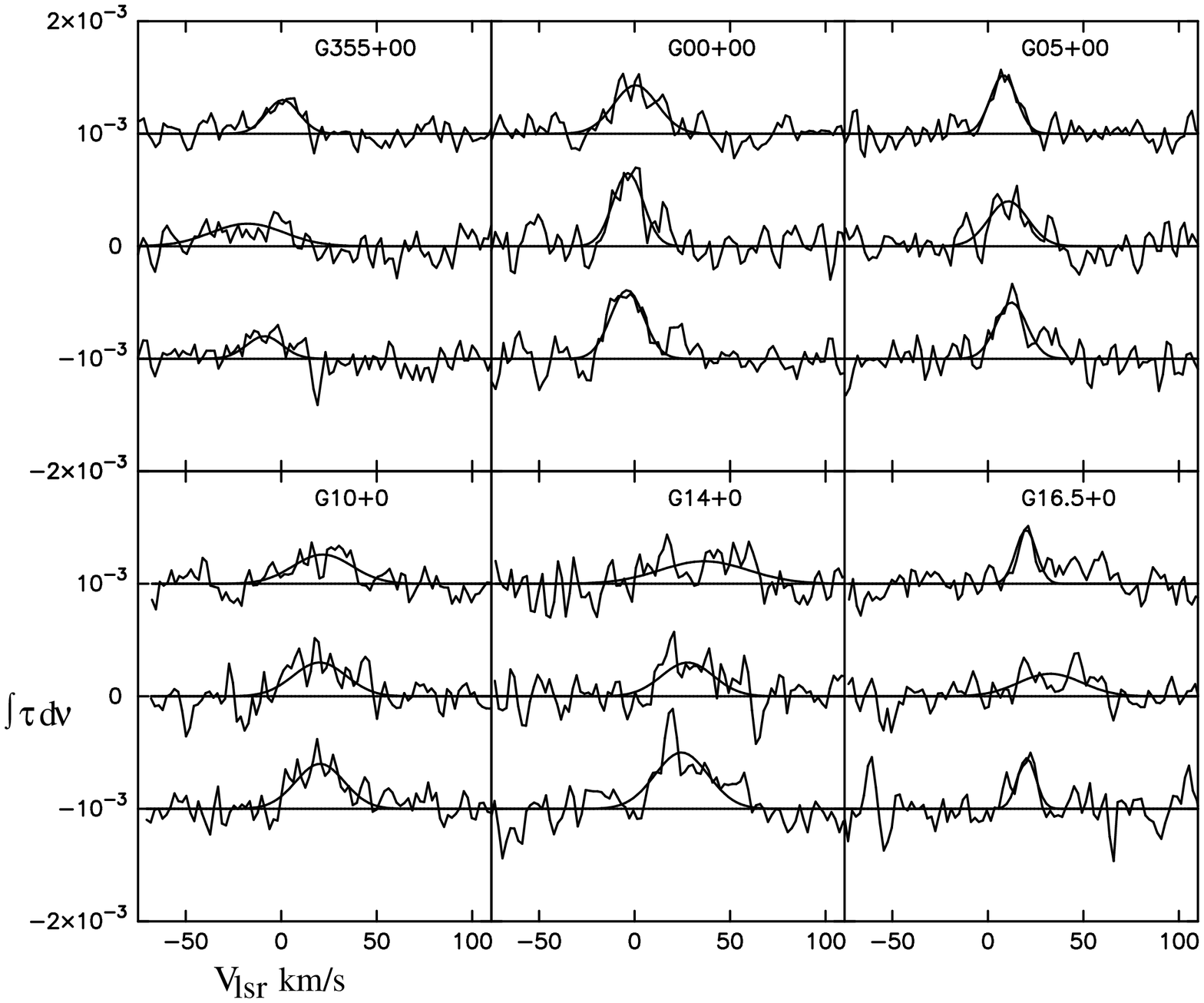,width=11cm,angle=-90}
\caption{\sf Spectra of carbon lines detected near 327 MHz using beam 5 and beam 7 of the
full ORT and using a single module of the ORT are shown.  In each frame, the lowermost
spectrum is from beam 5 and uppermost is from a single module.
These two spectra are offset along the y$-$axis from their zero$-$position.  The vertical
scale applies to all these spectra.  The jagged line is the observed spectrum
and the smooth line superposed on it is the Gaussian fit to the profile.}
\label{fig2}
\end{figure}

Totally, 12 positions (see Table \ref{tab3}) were observed near 328 MHz with the ORT.  
At the higher resolution, lines were detected from seven of these directions 
whereas in the low-resolution observations, 
lines were detected from six directions.  The line profiles are shown in
Fig \ref{fig2}.

\begin{table}
\begin{center}
\caption{\sf Parameters of the Carbon lines observed with the ORT.
 $\Delta V $ is the FWHM of the line.
The number in the brackets is the $1\sigma$ uncertainty.}

\vspace{0.5cm}
\begin{tabular}{lcccc}
\hline
{\bf Source} & $\bf T_{l}/T_{sys}$ & {\bf \Vlsr} & $\bf \Delta V $  & {\bf  \Itau} \\
  & $\times 10^{-3}$ & \kms & \kms & $s^{-1}$ \\
    & & & & \\
\multicolumn{5}{c}{\large \bf ORT Beam 7} \\
\hline
 G355+0 & 0.2(0.02) & -17.6(3.0) & 45.0(7.1) & 10.5(2.5)  \\
 G00+00 & 0.65(0.07) & -3.4(0.5)  & 18.7(1.1)  & 14.2(2.6) \\
 G05+00 &0.4(0.03) & 10.7(0.8) & 24.2(1.9) & 11.1(1.0)   \\
 G10+00 & 0.3(0.02)   & 20.1(1.3)  & 32.8(3.0)  & 11.1(1.2)  \\
 G14+00 & 0.3(0.02)   & 27.7(1.0) & 31.9(2.3) & 10.9(0.9)  \\
 G16.5+0& 0.2(0.02) & 32.7(2.3) & 39.1(5.3) & 9.1(2.3) \\
 G30+00 & $<$ 0.15  & - & 6.2 & $<1.1$   \\
 W49    & 0.3(0.06) & 42.2(1.5) & 30.7(3.6) & 10.7(2.4) \\
 W51    & $<$ 0.12 & -  & 6.2 & $<1.1$  \\
 G62+00 & $<$ 0.14 & -  & 6.2 & $<1.2$ \\
 G75+00 & $<$ 0.12  &-& 6.2  & $<1.1$  \\
 NGC2024 & $<$ 0.11  & - & 6.2  & $<1.1$ \\
\hline
    & & & & \\
\multicolumn{5}{c}{\large \bf ORT Beam 5} \\
\hline
 G355+0 & 0.2(0.04) & -8.8(1.9) & 21.9(4.4) & 5.1(2.5) \\
 G00+00 &  0.61(0.07)  & -4.1(0.5) & 20.6(1.1)  & 14.8(1.8) \\
 G05+00 & 0.5(0.03)   & 12.3(0.6)  & 19.0(1.4)  & 10.9(0.9) \\
 G10+00 & 0.4(0.02)  & 20.3(0.8) & 29.9(2.0) & 13.7(1.1) \\
 G14+00 & 0.5(0.06) & 24.6(1.7) & 32.7(3.9) & 18.8(1.2) \\
 G16.5+0& 0.45(0.1)  & 20.3(1.4) & 11.4(3.3) & 6.5(0.8) \\
 G30+00 &  $<$ 0.20& - & 6.2 & $<1.5$ \\
 W49    &  $<$ 0.17 & -  &  6.2  & $<1.2$ \\
 W51    & $<$ 0.14 & - & 6.2 &$<1.2$ \\
 G62+00 &  $<$ 0.13 & - & 6.2 & $<1.1$ \\
 G75+00 & $<$ 0.19  &- & 6.2 & $<1.4$ \\
 NGC2024 & $<$ 0.11  &- & 6.2 & $<1.1$ \\
\hline
    & & & & \\
\multicolumn{5}{c}{\large \bf ORT Single Module} \\
\hline
 G355+0 &  0.3(0.02)   & 0.8(0.7)  & 20.1(1.7)  & 6.9(0.7) \\
 G00+00 &  0.43(0.05)   &  0.5(0.6)  & 27.0(1.5) & 13.5(1.8) \\
 G05+00 &  0.52(0.02) & 8.2(0.4) & 16.2(0.9) & 9.8(0.6)\\
 G10+00 & 0.26(0.02) & 21.7(1.2)  & 36.3(2.9) & 10.7(1.0) \\
 G14+00 &  0.2(0.04) & 36.5(3.7) & 54.6(8.7) & 12.7(4.1) \\
 G16.5+0 & 0.48(0.07)  & 20.0(0.8)  & 12.7(2.0)  & 7.0(0.6)\\
 W49    &  $<$ 0.10  & -  & 6.2  & $<1.0$ \\
 G50+00 &  0.2(0.02) & 50.5(1.9) & 35.4(4.6) & 8.3(2.5) \\
 G62+00 & $<$0.42 &- & 8.0 & $<$3.7 \\
\hline

\end{tabular}

\label{tab4}
\end{center}
\end{table}

The observed spectra were modelled by Gaussian profiles and
the parameters are listed in Table \ref{tab4}.  
The carbon lines detected in the Galactic plane are broader (\delV $\ge$ 20 \kms)
and weaker as compared to those detected towards Cas A 
(\nocite{payne:89}Payne \et 1989) at this frequency. 
The lines detected using the full ORT and using the single module had comparable
strengths as is evident in Fig \ref{fig2}. 
No carbon line was detected towards other directions and the rms noise  
on the spectra are listed in Table \ref{tab3}.  

\subsection{VLA Observations at 332 MHz}

\begin{table}
\begin{center}
\caption{\sf Parameters of VLA Observations}
\vspace{0.3cm}
\begin{tabular}{ll}
\hline
Field Centre: $\alpha$(1950)  &$\rm 18^h12^m59^s$   \\
$~~~~~~~~~~~~~~~~~~~\delta$(1950)&$-16^{\circ}48'03''$  \\
 Observing Epoch   & June 1995    \\
 Duration of Observations       &  4.5 hours     \\
 Observed Transitions & 270$\alpha$, 271$\alpha$   \\
 Rest Frequencies (Carbon) & 332.419 MHz    \\
   &    328.76 MHz  \\
 Primary Beam    & 150'        \\
 Shortest spacing  & 0.035 km   \\
 Longest spacing   & 1.03 km    \\
 Observing Mode  & 4IF       \\
 Total bandwidth  & 781 kHz    \\ 
     &   (693 \kms)    \\
 Number of Channels   & 64     \\
 Frequency Resolution  & 12.207 kHz   \\
 Velocity Resolution   & 11 \kms   \\
 Amplitude Calibrator & 3C286    \\
 Phase Calibrator & 1827-360    \\
 Bandpass Calibrator & 1827-360    \\
 Synthesized beam  &321.4''$\times$209''  \\
    (natural weighting) & \\
 Rms noise, line   & 25 mJy/beam  \\
 Rms noise, continuum  & 150 mJy/beam  \\
\hline
\end{tabular}

\label{tab5}
\end{center}
\end{table}

Observations made with the ORT with two different angular resolutions, 
provided some preliminary information regarding the angular size of 
the line-forming region.  To obtain constraints on the possible
clumpiness of the line-forming gas, we used the VLA 
to observe the carbon line emission near 332 MHz ($270\alpha$) towards one 
of the positions in the Galactic plane which was detected with the ORT. 
The direction towards Galactic longitude $14^{\circ}$ in the Galactic plane was observed
for 4.5 hours in June 1995 using the D-configuration of the VLA.  
Details of this observation are given
in Table \ref{tab5}.  An amplitude calibrator was observed
at the start of the observations.  The phase calibrator which was
also used for bandpass calibration, was observed
once every 30 minutes.  The four-IF mode of the correlator which
consisted of two circular polarisations (Stokes RR and Stokes LL)
and two IF frequencies, was used.  Two recombination line
transitions ($270\alpha$ \& $271\alpha$) were observed simultaneously.

The UV data were processed using the standard procedures in the  
Astronomical Image Processing
System (AIPS) developed by NRAO.  The continuum image was generated 
by averaging the visibilities in the central three-quarters of the band. 
Natural weighting of the data gave a beam size of $5.4'\times3.5'$
with P.A.$=1.5^{\circ}$.  A line cube was obtained by Fourier 
transforming the residual visibilities after subtracting the continuum. 
To obtain the maximum possible signal-to-noise ratio, natural 
weighting was applied to the visibilities.
Although we aimed at achieving an rms noise of 7-8 mJy/beam
in the line images, severe problems due to interference forced us to discard more than 50 \%
of the data.  The rms noise on the line images that we 
finally obtained was 25 mJy/beam.

The $2^{\circ}$ field centred on $l=14^{\circ}, b=0^{\circ}$ which was imaged using 
the VLA at 332 MHz is shown in Fig \ref{fig3} (a). 
The rms noise in the image is 150 mJy/beam.  M17 (the bright source at 
the bottom-left corner of the
image), a \HII region, is the brightest source in the $2^{\circ}$ field with a 
peak brightness of 40 Jy/beam. 
The central regions of this nebula are optically thick at low frequencies. 
The observed brightness temperature at 330 MHz indicates an 
electron temperature of 7885 $^{\circ}$K for the peak emission of M17. 
which is comparable to the value 8000 K obtained by 
\nocite{subrahmanyan:96} Subrahmanyan \& Goss (1996).  Counterparts of other features 
in Fig \ref{fig3} (a) are identifiable on the 5 GHz \nocite{altenhoff:79}
(Altenhoff \et 1979) 
and 2.7 GHz (\nocite{reich:90}Reich \et 1990) continuum maps.

The line images showed no emission to a $3\sigma$ limit of 75 \mJyb. 
In Fig \ref{fig3} (b), the line emission integrated over the ORT beam
and divided by the integrated continuum flux (19 Jy) is shown.
No line emission is seen in the spectrum and the
upper limit on the line-to-continuum ratio of carbon and hydrogen lines
is $7.8\times10^{-3}$ ($3\sigma$ limit). 
This upper limit is consistent with the ORT results.  
The VLA observations place a lower limit of $10'$ on the size of `clumps', if any, in the
ORT beam.  This inference follows from a comparison of the results by
\nocite{anantha:85} Anantharamaiah(1985) who detected the C$272\alpha$ line with a peak flux of 160 mJy 
using a $2^{\circ}\times6'$ beam
and the $3\sigma$ detection limit of 75 mJy obtained from the present VLA data
with a beam of $5.4'\times 3.5'$.

\begin{figure}
\begin{center}
\epsfig{file=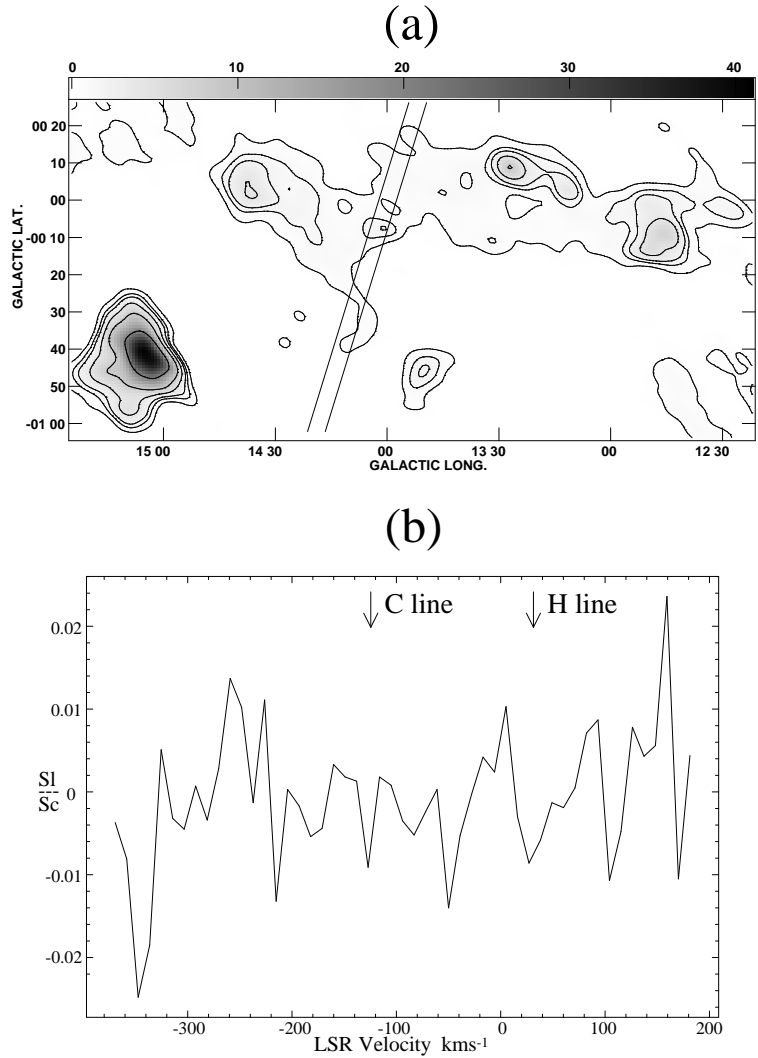,width=7cm,angle=-90} \\
\epsfig{file=fig3b.ps,width=7cm,angle=-90}
\caption{\sf (a) VLA Continuum Image of G14+00 at 330 MHz in Galactic coordinates.
The contour levels are 1,3,5,10,20,50,100,110 in units of
0.4 Jy/beam and the grey scale flux ranges
from 0 to 41 Jy/beam.  The full-ORT beam has been superposed on the
continuum map.  (b) Spectrum showing the line emission over the
full-ORT beam. The arrows show the expected positions of the carbon and hydrogen
recombination lines }
\label{fig3}
\vspace{5cm}
\end{center}
\end{figure}

\section{Constraints on the Physical Properties of the Carbon Line Regions}
\subsection{Angular Size of the Line forming Regions}
Since the interpretation of the observed recombination lines, in terms of the 
physical properties of the 
gas in which they arise, depends sensitively on the fraction of the beam that is
filled by the thermal gas, it is necessary to obtain a preliminary idea of its 
angular distribution.  
Our observations have been made at different angular resolutions ranging from
$21'\times 25^{\circ}$ at 34.5 MHz to $2^{\circ}\times2^{\circ}$ at 
328 MHz and we also make use of the line data at
76 MHz (\nocite{erickson:95} Erickson \et 1995) with a resolution of $4^{\circ}\times4^{\circ}$.  
Unless the line radiation fills the beam or
the distribution of the continuum intensity is highly concentrated and dominates 
the system temperature (\eg Cas~A),
beam dilution effects will be significant, and the actual line optical depths will be 
very different from the apparent (observed) optical depths.  
Comparison of the line-to-continuum ratios integrated over the width, 
observed with the two different
angular resolutions at 328 MHz (see Table \ref{tab4}) shows that, within
the errors, the two ratios are almost equal for most of the positions.  This equality
indicates that the angular extent of the cloud giving rise to these lines in the inner Galaxy 
is $\ge 2^{\circ}$ for
most lines of sight.  However, the possibility that the line-forming regions are 
made of a number of nearly
uniformly distributed clumps within the field of view is not ruled out 
by the ORT observations.
The VLA observations have placed a lower limit of $10'$ on the size of `clumps', 
if any, in the ORT beam.  
  
From their observation of carbon lines near 76 MHz as a function of Galactic latitude, 
\nocite{erickson:95} Erickson \et (1995)
concluded that the angular extent of the carbon line-forming regions in the inner Galaxy 
must be approximately $4^{\circ}$. 
Very low-frequency (near 25 MHz) observation towards $l = 75^{\circ}$
in the Galactic plane (\nocite{konovalenko:84a} Konovalenko 1984a) has also indicated that the 
carbon line-producing region is $\ge 4^{\circ}$ in extent.

Thus, from the existing observational data on low-frequency carbon 
recombination lines, it appears that the ionized carbon gas is 
distributed in form of clouds or clumps with individual sizes $>10'$ and
with an overall extent of at least a few degrees. 

\subsection{Constraints from the observed line widths}
One notable difference between the carbon lines observed towards Cas A 
and the lines observed from 
positions in the Galactic plane is the width of the lines (Table \ref{tab6}).  
While the lines detected towards Cas A (last row of Table \ref{tab6})
are seen to broaden with increase in quantum number, which is a clear
indication of pressure \& radiation broadening, no such trend is seen in
the lines observed from the Galactic plane.

The lack of $n-$dependent line broadening towards the Galactic plane seems to indicate
that lower electron densities and weaker ambient radiation fields 
prevail here compared to those in front of Cas A.  However,
the observed widths ($\ge 20$ \kms) of the lines from the Galactic plane 
cannot be explained by thermal motions and micro-turbulence ($\sim$ few \kms) 
in the cool clouds.  We suggest that the lines are broadened by the systematic 
motions owing to differential Galactic rotation and hence lack an $n-$dependence.

\begin{table}
\caption{\sf Widths of carbon recombination lines observed at different frequencies. }
\vspace{0.2cm}

\begin{tabular}{lcccc}
\hline
 & {\bf n$\sim$686} & {\bf n$\sim$575} & 
   {\bf n$\sim$443} & {\bf n$\sim$271} \\
{\bf Position} & {\bf 25 MHz } & {\bf 34.5 MHz } & 
                 {\bf 76 MHz} & {\bf 328 MHz} \\  
      & \kms & \kms & \kms & \kms\\
\hline
 G352+0  & - & $36.4(3.6)$ &  $11(1)$ & $20.1(1.7)$ \\ 
        &    &          &         & (G355+00) \\
 G00+00 & - & $20.5(1.2)$  & $24(1)$ & $27.0(1.5)$ \\ 
 G05+00 & - & $21.0(2.5)$  & $25(4)$ & $16.2(0.9)$ \\ 
        &   &  &           (G06+00) &\\
 G10+00 & - & $37.0(3.2)$ & $26(2)$ & $36.3(2.9)$ \\ 
 G14+00 & - & $56.0(4.5)$  & $25(3)$  & $54.6(8.7)$ \\ 
 G16.5+0 & - &  $32.6(3.4)$  & $47(4)$ & $12.7(2)$ \\ 
 G63+00 & - & $45.9(4.4)$  &- & - \\ 
 G75+00 & $15(0.9)$\footnote{from Konovalenko (1984a)} 
  & $24.4(2.8)$  & - & -  \\ 
 DR21   & $42(12)$\footnote{from Golynkin and Konovalenko (1991)}
         & $18.5(2.7)$  & - & -\\
 Cas A  & $71.9(16.4)$\footnote{from Konovalenko (1984b)} 
         & $26.0(3.1)$  
       & $6.7(0.4)$\footnote{from PAE89} & $5.0(0.5)$ \\ 

\hline
\end{tabular}
\\
\small
{$^1$ from Konovalenko (1984a)\\}
{$^2$ from Golynkin and Konovalenko (1991)\\}
{$^3$ from Konovalenko (1984b)\\}
{$^4$ from PAE89\\}


\label{tab6}
\end{table}

Since pressure($\Delta V_P$) and radiation($\Delta V_R$) broadenings are expected to be maximum at
34.5 MHz, the line widths at this frequency can be used to derive upper 
limits on the electron density and the ambient radiation field.
We used Eqns \ref{eq:Press2} and \ref{eq:Rad} from the paper by 
\nocite{shaver:75}Shaver (1975) 
for deriving the limits which are listed in Table \ref{tab7}; and
we reproduce them below for ready reference
\beqn
\Delta V_P = 2 \times 10^{-8}~exp \left( -\frac{26}{T_e^{1/3}} \right)~\frac{n_e n^{5.2}}{T_e^{1.5}}
~\frac{c/(kms^{-1})}{ \nu/(kHz)} \rm ~~~ kms^{-1}
\label{eq:Press2}
\eeqn
\beqn
\Delta V_R = 8 \times 10^{-20} ~W_\nu ~T_{R,100} ~n^{5.8} ~ \frac{c/(kms^{-1})}{\nu/(kHz)} ~~~ {\rm kms^{-1}} ~
\label{eq:Rad}
\eeqn

where $\Delta V_P$ and $\Delta V_R$ are the FWHM due to pressure and radiation broadening
(kms$^-1$),
$T_e$ is the electron temperature (K), $n_e$ is the electron density (\cm3 ), $n$
is the quantum number, $c$ 
is the velocity of light (kms$^{-1}$), $\nu$ is the frequency (kHz), $W_\nu$ is the dilution
factor, $T_{R,100}$ is the radiation temperature (K) at 100 MHz.

The widths listed in Table \ref{tab7} were calculated assuming that the
entire width is due to either of the
broadening mechanism.  The upper limits on the electron density in 
the medium were calculated for a nebula at 20 K.  

\begin{table}
\caption{\sf Upper limits on the electron density and radiation temperature
(at 100 MHz) from the observations at 34.5 MHz}
\vspace{0.5cm}
\begin{center}
\begin{tabular}{lccccc}
\hline

%
{\bf Position}&{\bf \delV$_{obs}$} & $\bf n_{e,max}^{1}$ &
              $\bf T_{R100,max}^{1}$ & $\bf T_{R100}^{2}$  &
                \delV$_{exp}^{3}$  \\
  & \kms  & \cm3 & K & K & \kms \\
\hline
 G352+00 & 36.4(3.6) &  1.21(0.12) & 5160(510) & 3460 & 24.4 \\ 
 G00+00  & 20.5(1.2) &  0.68(0.04) & 2906(170) & 5030  & 35.5 \\
 G05+00  & 21.0(2.5) &  0.7(0.08)  & 2977(354) & 5030 & 35.5 \\ 
 G10+00  & 37.0(3.2) &  1.23(0.11) & 5245(454) & 5660 & 40.0 \\ 
 G14+00  & 56.0(4.5) &  1.86(0.15) & 7938(638) & 5660 & 40.0 \\
 G16.5+0 & 32.6(3.4) &  1.08(0.11) & 4621(482) & 5030 & 35.5 \\
 G63+00  & 45.9(4.4) &  1.51(0.15) & 6507(624) & 2830 & 20.0 \\
 G75+00  & 24.4(2.8) &  0.81(0.09) & 3459(370) & 2830 & 20.0 \\
 DR21    & 18.5(2.7) &  0.61(0.09) & 2622(383) & 2680 & 19.0 \\ 
\hline
\end{tabular}
\end{center}

\small
{$^1$ Upper limit derived from the observed line width
at 34.5 MHz and using \Te $=20$ K and $W_\nu=1$. \\
$^2$ Averaged value over the Gauribidanur beam of $25^{\circ}$
obtained from the continuum map at 34.5 MHz (Dwarakanath 1989)
and extrapolated to 100 MHz using a spectral index $\alpha= 2.6$ where
$T_B \propto \nu^{-\alpha}$. \\
$^3$ Expected line broadening at 34.5 MHz due to \Tr noted
in column 6 and assuming $W_\nu=1$. \\}


\label{tab7}
\end{table}

The insignificant contribution of pressure and radiation 
broadening to the line width at 34.5 MHz indicates that the actual electron 
densities and radiation fields are 
much smaller than those listed in columns 4 and 5 of Table \ref{tab7}.  
Furthermore, the electron density decreases if the kinetic temperature 
is increased $e.g.$ the values of \ne noted in
column 4 will decrease by a factor of $\sim 4$ if \Te increased to 75 K. 
The derived upper limit for the radiation temperature in the Table assumes a dilution
factor of unity which implies that the absorbing cloud is 
isotropically illuminated by the non-thermal radiation field.  The actual radiation temperature
seen by the cloud could be equal to or less than the values listed in column 6, 
which were obtained from the $26'\times42'$ continuum map of 
\nocite{dwarkanath:89} Dwarakanath (1989) at 34.5 MHz.
The widths of the carbon lines near 34.5 MHz as predicted by this 
radiation field are listed in column 7.  

The limits listed in Table \ref{tab7} are useful because they 
define the absolute boundaries of
the parameter space that we use for modelling the observed recombination lines. 

\subsection{Longitude--Velocity diagram} 
The l-v diagram for the carbon recombination lines observed at 34.5 and 328 MHz is
shown in \mbox{Fig \ref{fig4}}.  The Galactic rotation model by
\nocite{burton:88} Burton (1988) has been
used.  The kinematic nature displayed by gas at Galactocentric distances of
7.5 kpc (solid line) and 4 kpc (dashed line) is also shown in the figure.

\begin{figure}
\mbox{\epsfig{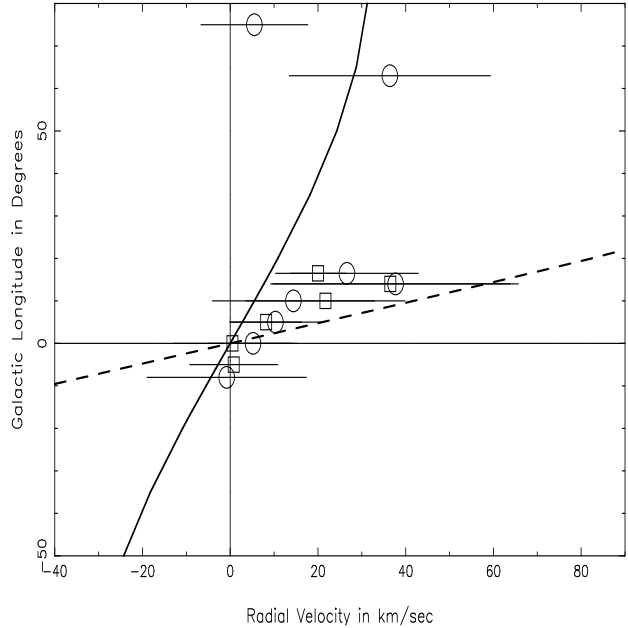}}
\caption{\sf The Longitude-Velocity diagram of the carbon
recombination lines at 34.5 MHz(circles) and 328 MHz(squares) from
the Galactic plane.  The horizontal lines about each point represent the
observed widths of the lines.  The solid line shows the variation
of the radial velocity with longitude for a cloud placed at
a Galactocentric distance of 7.5 kpc whereas the dashed line
shows the expected l-v plot for a cloud at a distance of 4 kpc.
These are obtained using the Galactic rotation model of Burton (1988).}
\label{fig4}
\end{figure}

All the observed points with $l > 0^{\circ}$ lie between 4 and 7.5 kpc.  
This range coincides with the regions observe to be rich in molecular clouds
(\nocite{scoville:75} Scoville \& Solomon 1975), H$166\alpha$ 
emission (\nocite{lockman:76} Lockman 1976) and 
strong \HII regions (\nocite{downes:80}Downes \et 1980).   Broad lines of
\HI emission (\nocite{burton:88}Burton 1988) and absorption 
(\nocite{garwood:89}Garwood \& Dickey 1989), 
$^{12}$CO (\nocite{bania:77}Bania, 1977), H$166\alpha$ (\nocite{lockman:76} Lockman 1976) and
[\CII ] 158 $\mu$m (\nocite{mizutani:94}Mizutani \et 1994) emission are 
observed towards the Galactic centre.
These broad ($> 100$ \kms) lines are due to
non-circular motions near the Galactic centre.  However, the low-frequency carbon
recombination lines observed by us are relatively narrow ($\sim 20$ \kms).  
The hydrogen recombination lines near 327 MHz observed by
\nocite{anantha:85} Anantharamaiah (1985) and \nocite{roshi:97}Roshi \& Anantharamaiah (1997) 
are also relatively narrow.
Excluding the Galactic centre direction, the
l-v distribution of the carbon lines that we observe resembles 
that traced by \HI absorption, 
diffuse \HII (\nocite{lockman:96}Lockman \et 1996) and $^{12}$CO (except for the features 
due to the 3 kpc arm).\nocite{erickson:95}  
Erickson \et (1995) have a more continuous sampling of carbon lines
near 76 MHz ($n \sim 445$) in the inner Galaxy and they also find that the gas responsible
for the lines lies between galactocentric distances of 5 and 8 kpc.  With the available
data, we can only say that the gas giving rise to carbon lines 
in the inner Galaxy is likely to be distributed
between galactocentric distances of 4 and 8 kpc and is likely to be associated with
either the cold \HI gas, or the molecular gas or the  \HII regions.  
From the similarity of the $l-v$ distributions
of the emission lines observed near 
328 MHz and the absorption lines detected near 34.5 MHz and 76 MHz 
from the inner Galaxy ($l<17^{\circ}$), we assume that
all these lines arise in the same ionized carbon gas.

\subsection{Origin of carbon lines in cold, neutral gas}
Although the widths of the observed lines do not rule out an origin in 
classical \HII regions, the
low electron densities (previous section) and the absence of hydrogen recombination
lines at low frequencies
rule out such an origin.  Hence, the lines are most likely associated
with cold gas where only carbon is ionized.  The possiblities are either
the atomic \HI gas (the cold neutral medium), the molecular gas or the low-excitation
photodissociation regions in the interstellar medium. 
Interpretation of low-frequency recombination lines observed towards the direction of Cas A 
has shown that these lines are most likely to be associated with atomic \HI gas.
However, the different pressure and radiation broadening seen in the low-frequency
lines observed from Cas A and 
the Galactic plane suggests that detectable low-frequency lines may arise in 
regions with a range of physical properties.  In a review paper, 
\nocite{sorochenko:96} Sorochenko (1996) discusses
Cas A and other such directions towards which carbon lines have been detected
and concludes that \CII regions are formed on the 
surface of molecular clouds exposed to external
ionizing UV radiation either close to \HII regions or around isolated molecular clouds.

\section{Modelling the line-forming regions }
\subsection{Assumptions of the model}
We assumed a cloud of uniform density \ne, temperature \Te and emission measure
EM illuminated by a non-thermal radiation field characterized by a brightness 
temperature \Tr at 100 MHz.  
Although this is a simplified model and the real cloud is likely to
contain density condensations, it can be 
used to derive the average properties of the medium under consideration.  
The solution of the radiative transfer equation is
(\nocite{shaver:75}Shaver 1975): 
\beqn
T_l = T_{\circ}\left[e^{-\tau_c}(e^{-b_n\beta_n\tau_l^*} - 1)\right] ~~
+ ~~T_e \left[ \frac{(b_n\tau_l^* + \tau_c)}{ \tau_l+\tau_c}
( 1 - e^{-\tau_l+\tau_c}) - (1- e^{-\tau_c}) \right] 
\label{eq:Tl1}
\eeqn 
In Eqn \ref{eq:Tl1}, $T_l$ is the line temperature, $T_{\circ}$ is the background temperature, 
\Te is the electron temperature, 
$\tau_c$ is the continuum optical depth at the given frequency, $\tau_l^*$ is the LTE
line optical depth, $\tau_l^*$ is the non-LTE line optical depth 
and \bn and $\beta_n$ are the departure coefficients 
which measure the deviation of the level populations from LTE values.  Further 
simplifying assumptions which approximate the real system are made.  At low frequencies, 
the lines are sensitive to tenous, cool nebulae \ie \Te $<< T_{\circ}$.  
Since $\tau_c, \tau_l << 1$   
at these frequencies, the second term in the above equation 
is much smaller than the first term
and can be neglected.  Hence, the line temperature can be 
expressed under these conditions as:

\beqn
T_l = T_{\circ}\left[e^{-\tau_c}(e^{-b_n\beta_n\tau_l^*} - 1)\right]
\eeqn
Since $T_c = T_{\circ}~e^{-\tau_c}$ and $\tau_c << 1$, the line-to-continuum ratio 
can be approximated as
\beqn
\frac{T_l}{T_c} = -b_n\beta_n\tau_l^* = -\tau_l
\label{Eqn:taul}
\eeqn
The departure coefficients \bn and $\beta_n$ quantify the non-LTE effects 
influencing the level populations.  The true population of 
a level $n$ is given as $N_n = b_nN_n^*$ where 
$N_n^*$  is the expected LTE population. 
$\beta_n = 1 - \frac{kT_e}{h\nu} \frac{d(lnb_n)}{dn}$ depends on the gradient of \bn with $n$.
A negative value of $\beta_n$ signifies inverted populations and hence 
stimulated emission.  A positive value of $\beta_n$ means that the 
line at that frequency is likely to appear 
in absorption if the background temperature $T_{\circ} > T_e$.  
The RHS of Eqn \ref{Eqn:taul} is the non-LTE value of the line optical depth.  

Rewriting Eqn \ref{Eqn:taul} in terms of antenna temperatures:
\beqn
\tau_l = \frac{\Omega_c}{\Omega_L}~~\frac{T_{A,l}}{T_{A,c}}
\label{Eq7}
\eeqn
where $\Omega_c, \Omega_l$ are the solid angles subtended by the continuum and the 
line-forming regions.  $\Omega_c, \Omega_l \le \Omega_B$ where $\Omega_B$ is the 
solid angle subtended by the telescope beam.  $T_{A,l}$ and $T_{A,c}$ are the
line and continuum antenna temperatures.  At 34.5 MHz, 
the non-thermal background radiation dominates the continuum temperature.  
Since this background is all-pervasive in the Galaxy,
$\Omega_c \ge \Omega_B$ and the main factor responsible for the unknown beam 
dilution factor is the extent of the line-forming gas within the observing beam.  

At low frequencies, line-widths are likely to be affected by various $n$-dependent 
line broadening mechanisms, leading to reduction
in the peak optical depths and increased widths.  Therefore, the integrated optical
depth is the correct physical quantity to compare at different frequencies.  
Theoretically, the integral of the line optical depth can be given by
\nocite{payne:94} Payne \et (1994):

\beqn
\int\tau_l d\nu = 2.046\times10^6~~T_e^{-5/2}e^{\chi_n}~EM_l~b_n\beta_n ~~~s^{-1}
\label{eqn:Tl7}
\eeqn
 
Using this Eqn, we attempt to fit the observed variation in the integrated optical depth
(given by Eqn \ref{Eqn:taul}) with principal quantum 
number for plausible combinations of \Te and \ne . 

\subsection{The Modelling Procedure}
We use carbon line observations at three frequencies to model the low-excitation \CII regions
- our observations at 34.5 MHz and 328 MHz and the 
Parkes observations by \nocite{erickson:95} Erickson \et (1995) at 76 MHz.  
The angular resolution of these observations are different.  
We assume that the lines at the three frequencies arise in the same gas.

Each model is characterized by an electron temperature $T_e$, electron density $n_e$,
emission measure EM, radiation temperature \Tr, carbon depletion factor $\delta_C$ and the
angular size of the line-forming region.  As discussed earlier, 
the upper limits on \Tr and \ne were determined from the widths of the 
lines observed at 34.5 MHz.
We consider a range of possible temperatures for the line-forming region 
which include the typical temperatures of cold \HI, $\rm H_2$, and photo-dissociation regions 
which are all probable places of origin for the carbon lines.  
The lower limit on the electron densities is set from the maximum pathlengths
through the Galaxy.  We considered models with a carbon depletion factor of 0.5. 
Since observations indicate that the line-forming region could have an angular size 
$\ge 2^{\circ}$,  we explore models in which the line-forming region is 
(a) ubiquitous (\ie fills all the beams that are used); (b) $4^{\circ}$ and (c) $2^{\circ}$. 
The departure coefficients $b_n$ and $\beta_n$ which account for the non-LTE 
effects were calculated using the computer code, originally due 
to \nocite{salem:79} Salem \& Brocklehurst (1979)
and later modified by \nocite{walmsley:82} Walmsley \& Watson (1982) and 
\nocite{payne:94} Payne \et (1994).

The model integrated optical depths are 
calculated for different $n$ using Eqn~\ref{eqn:Tl7}
and the models are normalized to the observed integrated optical depth at 328 MHz. 
The ratio of line temperature to system temperature is considered to be equivalent to 
the line-to-continuum ratio, since the system temperature is dominated
by the strong sky background at these low frequencies. 
Any model that we generate needs to satisfy the observed
integrated optical depth \Itau and be consistent with the upper limits
on the physical conditions placed by the line data at 34.5 MHz.
Since the observed width of the line is independent of frequency, we assumed
that the contribution to the width at 34.5 MHz due to pressure broadening
is $< 5$ \kms.  This assumption translated to an upper limit
on the electron density of 0.3 \cm3 .  The lower limit was 0.001 \cm3 .
The electron temperature space was varied from 20~K to 400~K.  
These models were generated for two values of the radiation temperature
at 100~MHz which were 1250~K and 2500~K. 

The non-LTE population is calculated assuming that the quantum levels are in
statistical equilibrium so that the rate of population of a level n due to all 
possible physical processes is equal to the rate of depopulation of that level
(\nocite{shaver:75}Shaver 1975).  The main physical processes which
determine the level populations are collisional and radiative excitation and 
de-excitation.  The departure coefficients are calculated by solving
a series of differential equations after imposing a boundary 
condition \SB79 (\nocite{salem:79} Salem \& Brocklehurst 1979) 
\ie the levels are assumed to be in
LTE at very large $n$.  
In case of carbon, it is necessary to also consider a 
dielectronic-like recombination process which significantly
influences the high-$n$ populations if the temperatures are around 100 K
(\nocite{watson:80} Watson \et 1980) 

\subsection{Results of modelling} 
We examine the models which explain the observed variation in the
integrated optical depth with frequency. 
The criteria used to select the possible physical models are
(a) the observed behaviour of the integrated optical depth as a function of $n$,
(b) the path lengths through the region which should be $\le 5$ kpc, as required
by the observed line widths and  
(c) the turnover from emission to absorption which should occur at $n < 443$ 
as the lines are observed in absorption at 76 MHz (\nocite{erickson:95} Erickson \et 1995).
We classify the results on the basis of the assumed angular size of 
the line-forming regions.  The angular size determines the beam
dilution and the scaling from the observed to true optical depth.  The data
at the three frequencies have different angular resolutions and therefore 
different beam-filling factors.  In all the cases, we assume, for simplicity,
that the non-thermal radiation field uniformly fills the beams.

\begin{center}
\begin{table}
\caption{\sf Model parameters for three positions for three assumed cloud sizes.
\Tr = 1250 K.}
\vspace{0.3cm}
\begin{center}
\begin{tabular}{ccccccc}
\hline
{\large \bf Position } & $\large \bf T_e$  & {$\large \bf n_e$} & {$ \large \bf EM$} &  {$\large \bf Size$} & {$\large \bf n_HT_e$} &  \\
& $ \bf K$  & {$ \bf cm^{-3}$} & {$ \bf pc cm^{-6}$} &  {$ \bf pc$} & 
      {$\bf cm^{-3} K$} &  \\
& & & & & &\\
\multicolumn{7}{c}{$\large \bf (a)~~ Cloud size > 4^{\circ}$} \\
\hline
G00+00 & 40 & 0.1 & 0.009 & 0.9 & 26680 & \\ 
       & 20 & 0.3 & 0.006 & 0.07 & 40000 &  \\
& & & & & &\\
G05+00 & 20 & 0.1 & 0.002 & 0.2 & 13340 & \\
       & 40 & 0.1 & 0.007 & 0.7 & 26680 & \\
& & & & & &\\
G14+00 & 40 & 0.1 & 0.007 & 0.7 & 26680 & \\
\hline
& & & & & &\\
& & & & & & \\
\multicolumn{7}{c}{$\large \bf (b)~~ Cloud size = 4^{\circ}$} \\
\hline
G00+00 & 80 & 0.03 & 0.019 & 21.1 & 16000 & \\
       & 200 & 0.03 & 0.079  & 87.8  & 40000& \\
& & & & & & \\
G05+00 & 60 & 0.05 & 0.01 & 4.0  & 19980 & \\
       & 80 & 0.03  & 0.014 & 15.6 & 16000 & \\
& & & & & & \\
G14+00 & 80 & 0.03 & 0.018 & 20.0 & 16000 &  \\
       & 100  & 0.03  & 0.025 & 27.8 & 20000 & \\
\hline
& & & & & &\\
& & & & & &\\
\multicolumn{7}{c}{$\large \bf (c)~~ Cloud size = 2^{\circ}$} \\
\hline
G00+00 & 150 & 0.03 & 0.051 & 56.7 & 30000& \\
       & 300 & 0.03 & 0.151 & 167.8 & 60000    & \\
& & & & & &\\
G05+00 & 150 & 0.03 & 0.037 & 41.1  & 30000  & \\
       & 200  & 0.03  & 0.058 & 64.4 & 40000 & \\
& & & & & &\\
G14+00 & 200  & 0.03 & 0.075 & 83.3 & 40000 & \\
       & 300 & 0.03 & 0.142  & 157.8  & 60000 & \\
\hline
\end{tabular}
\end{center}

\label{tab8}
\end{table}
\end{center}

\subsubsection{Case 1: Cloud size $ > 4^{\circ}$ or equal beam coverage by 
line and continuum:} 
We found that remarkably similar physical models reasonably explained 
the observed data towards
the six positions in the inner Galaxy.  The best-fitting
models towards three of the positions, $l=0^{\circ}$, $l=5^{\circ}$, and 
$l=14^{\circ}$ are shown in Fig \ref{fig5} (panels on left) and the parameters are 
listed in Table \ref{tab8}(a).  In all these models, we assumed
\Tr $=1250$ K.  Similar physical properties were found if 
\Tr $=2500$ K.  All the models were constrained by the data at 328 MHz. 
The turnover from emission to absorption occurs around
$n\sim 400$.  We found no model which simultaneously
explains the data at 76 MHz and 34.5 MHz, which is most likely due to the
unknown beam dilution at 34.5 MHz.  The models which explained the observed
variation in \Itau with $n$ were within the temperature range $20-40$ K.  
Outside this range, the models predicted large optical depths at 76 MHz and 34.5 MHz.
The electron densities in the acceptable models are 0.1 \cm3 .  Although higher electron densities
provided better fits to the data (e.g. the dashed line in Fig \ref{fig5}
for G00+00 has \ne = 0.3 \cm3 ),the
collisional broadening at 34.5 MHz exceeded the limit of 5 \kms and hence 
are not favoured.

The path lengths through the cloud range from 0.07 pc to 0.9 pc suggesting
a sheet-like morphology for the regions
and the emission measures are $\le 0.01$ pc $\rm cm^{-6}$.  Typical pressures in terms of
$n_HT_e$ are a few tens of thousands \cm3 K, which is higher than the average
thermal pressure in the interstellar medium.  We did not find any model which was
in pressure equilibrium with \HI.  The relatively low temperatures in these models suggest
association with molecular clouds.   

\begin{figure}
\vspace{4cm}
\epsfig{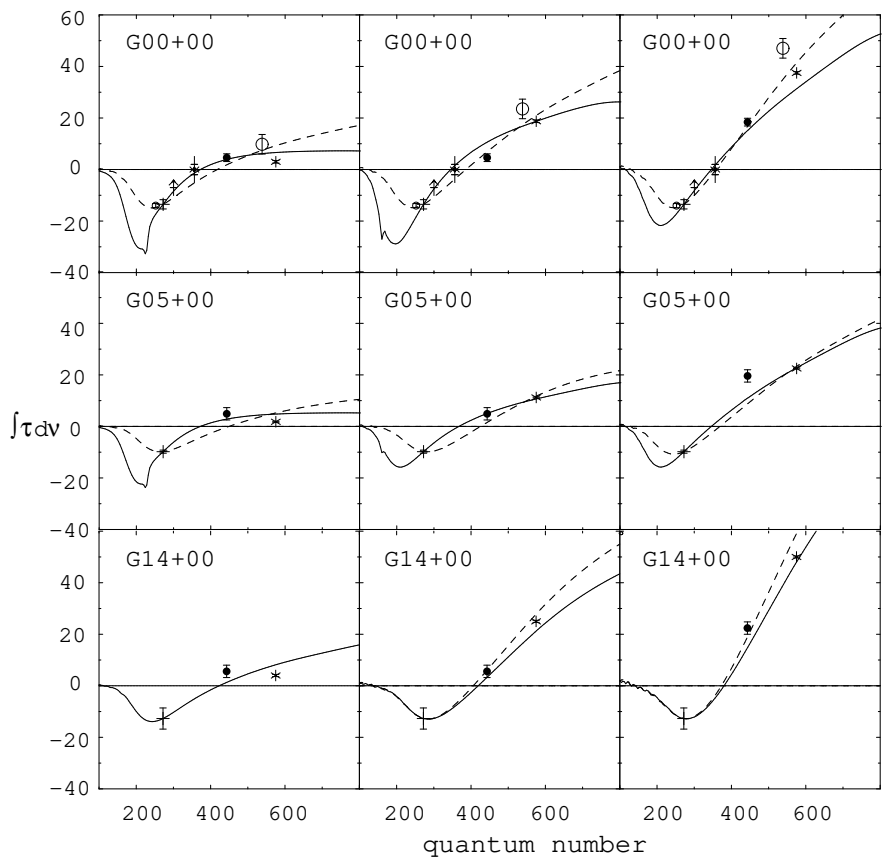}
\vspace{-5cm}
\caption{\sf Model fits to observed carbon line data shown for three Galactic plane
positions.  The unbroken line and broken lines represent the two
best-fit models for the two sets of parameters given in Table \ref{tab8}.
The left panels
show models for clouds assumed to be $\ge 4^{\circ}$; the centre panels show model fits
for clouds assumed to be $=4^{\circ}$; and
the panels on the right show models for clouds $=2^{\circ}$.  Towards G00+00,
the $n=252, n=300$
points are from  Pedlar \et (1978),
$n=356$ is from Anantharamaiah \et (1988) and $n=530$
is from Smirnov \et (1996). }
\label{fig5}
\end{figure}

\subsubsection{Case 2: Cloud Size $ \bf = 4 ^{\circ}$} These regions, if placed at a 
radial distance of
3.5 kpc from us, would have a linear size of 250 pc which is comparable
to the scale height of the atomic gas.  The best-fitting physical models 
have \Te in the range of $60-80$ K, \ne in the range of $0.03-0.05$ \cm3 , and
path lengths range from 4 pc to 30 pc.  
Fig \ref{fig5} (centre panels) shows the models for the positions
$l=0^{\circ}$, $l=5^{\circ}$ and $l=14^{\circ}$ and
Table \ref{tab8}(b) lists the parameters.  All the models are for
\Tr $=1250$ K.  
The turnover from emission to absorption occurs around $n \sim 400$. 
The relatively warmer temperatures in these models favour association with atomic
\HI gas in the Galaxy.  However, the thermal pressures are more than 3 times 
the average interstellar pressure ($n_HT_e \sim 5000$ \cm3 K).  
There was no change in model parameters with \Tr $=2500$ K except for a slight  
lowering of quantum number at which the turnover occurs. 

\subsubsection{Case 3: Cloud Size $= 2^{\circ}$}  If these regions are placed
at a radial distance of 3.5 kpc, then their linear size would be $\sim 125$ pc.
Models for three Galactic plane positions are shown in Fig \ref{fig5} (panels on the right)
and the parameters are listed in Table \ref{tab8}(c).
Best-fitting models have \Te between 150 K and
200 K and the electron density \ne $\sim 0.03$ \cm3 .  Lower temperatures
are possible if the upper limit on the electron density is
relaxed i.e. the pressure broadening is $> 5$ \kms.  
The path lengths required by these models ranged from 40 to 170 pc.
The models with \Tr $=2500$ K slightly lowered the quantum
numbers at which the turnover occurs, but the overall model parameters remained 
similar.

\section{Discussion}
The modelling described in the 
previous sections showed that most of the positions within $l=352^{\circ}$ to
$l=17^{\circ}$ possess remarkably similar physical properties.  
Quantitatively, if the clouds along the line-of-sight are 
$> 4^{\circ}$ in extent, then typically
\Te $= 20-40$ K, $n_e \sim 0.1$ \cm3 ~and the pathlengths are tiny ($< 1$ pc).
On the other hand, if the cloud is $4^{\circ}$ in size then
it is likely to be at a temperature between $60-200$ K, with electron density 
between $\sim 0.03-0.05$ \cm3 . 
The path lengths through these clouds are in the range $4-90$ pc.  If
the cloud subtends an angle of $2^{\circ}$, then they are likely to have
\Te $=150-300$ K and \ne $=0.03$ \cm3 ~and pathlengths between 40 and 170 pc.
All these models are able to reasonably explain the observed variation in the
optical depth with $n$.  Since the data is limited and the angular resolution
is coarse, it
is not possible to derive more stringent constraints on the parameters.
It is entirely possible that the observed carbon recombination lines arise in 
clouds possessing a range of temperatures, electron
densities and angular sizes.  Here we discuss the plausibility of the 
above models using qualitative arguments.

The low temperature (\Te $= 20$ K) models give a good fit to the 
observed line strengths at two frequencies for most positions.  If these
models correctly explain the physical properties of the line-forming regions then
their inability to explain the third data point is probably telling us about
the beam dilution at the lowest resolution.  
However, these models require extremely short pathlengths
through the ionized-carbon gas.  Since the widths of the lines 
detected at frequencies
differing by a factor of $\sim 10$, are found to be very similar, 
it implies that pressure and radiation broadening, which are strong 
functions of $n$ is not responsible for the
line widths even at the lowest frequency (\ie $\sim 34.5$ MHz).  The 
origin of the observed width is, most likely, due to the differential Galactic rotation
with the gas distributed between galactocentric distances of 4 and 7 kpc.
However, the pathlengths required by the low-temperature models are tiny
$ < 1$ pc.  If this pathlength is to be divided into smaller regions 
and distributed over $\sim 3$ kpc (to explain the line width), 
then it would require individual clouds to 
be $ < 0.1$ pc thick which implies
an extremely thin sheet-like geometry since the angular extent of the 
clouds are $ > 4^{\circ}$.  This peculiar morphology for the clouds make the
models difficult to accept; however they are not entirely implausible.
These thin sheet-like structures are reminiscent of ionization fronts near 
star-forming regions which typically have a thickness of $\sim 0.02$ pc
suggesting a parallel scenario where carbon is photo-ionized in 
a thin outer layer of molecular clouds by the ambient ultraviolet radiation field.  
However, to expect many such sheet$-$like objects 
to be distributed along every line-of-sight
to generate the observed strength and widths appears too contrived.
In the case of Cas~A (the most-studied direction in recombination lines at low frequencies;
\nocite{payne:94} Payne \et, 1994
and the references therein), the low$-$\Te models were ruled out 
since they failed to explain the observed variation in optical 
depth.  
Another interesting difference is that
the carbon lines towards Cas~A were distinctly affected
by frequency-dependent line-broadening whereas the lines from the Galactic
plane positions hardly show any $n-$dependent broadening suggesting different
physical conditions, especially the electron density and radiation field.
The main drawback of the data we use here is that they are at only three frequencies,
and with different and coarse angular resolutions which makes beam dilution
an important issue.  In contrast, for the data towards Cas~A,
the beam-size was determined by the continuum emission from
Cas~A since it dominates the system noise.  
One argument in favour of the low-temperature models
is that the constraint 
on pressure equilibrium with the interstellar medium is lifted since the gas is assumed to
coexist with molecular gas.  Moreover, since the carbon lines seem to be widely
detectable in the inner Galaxy, it also favours association 
with molecular gas which has maximum surface brightness in that 
region.  

The higher temperature (\Te $\ge 60$ K) models, which indirectly 
assume coexistence of ionized carbon with atomic \HI gas in the Galaxy, 
successfully fit the observed data towards most of the inner Galaxy
positions if the angular size of the clouds is $4^{\circ}$ or $2^{\circ}$.  
Due to the low electron densities ($\sim 0.03$ \cm3 ) in these models, 
they are sensitive to changes in the radiation temperature.  
For \Tr $=5500$ K (such high radiation temperatures are 
ruled out by the observed line widths 
near 34.5 MHz), data from all the positions could be fitted assuming a 
cloud size of $4^{\circ}$ or $2^{\circ}$, whereas for \Tr $=1250$ K 
(which produces a radiation width of $\le 5$ \kms at 34.5 MHz for a dilution
factor of 0.5), a few of the positions could not be fitted to the model.
With the present data, we cannot obtain a more
quantitative interpretation of this result.  As mentioned before, the 
range of physical conditions in the clouds predicted by the warm models appears to be 
well-constrained.  However, we had to relax the criterion of 
pressure equilibrium with \HI to obtain good fits to
most of the data.  So, all the models in which the temperature favours association
with \HI have kinetic pressures which range from $3-10$ times the interstellar pressures.
These models do predict reasonable pathlengths for the line-forming clouds.    
Recall that towards Cas~A, the warm models 
(\nocite{payne:89} Payne \et 1989, \nocite{payne:94} Payne \et 1994) satisfactorily explained the
observed variation in optical depth over a wide frequency range. 
Thus, we favour the view that a significant fraction of low-frequency 
carbon recombination lines arise in warm \HI gas 
in the photodissociation regions.

The range of physical parameters that we obtain in this 
analysis seems to favour an association
of the \CII gas with the photo-dissociation regions. 
A photo-dissociation region as defined by \nocite{tielens:85} Tielens \& 
Hollenbach (1985) are ``regions where
FUV radiation ($6-13.6$ eV) dominates the heating and/or some 
important aspect of the chemistry'' and contains most of the atomic and 
molecular gas in the Galaxy.  The low-excitation photo-dissociation region discussed
by \nocite{hollenbach:91} Hollenbach \et (1991) are illuminated by the interstellar radiation field.  
These regions, according to them, include a warm (T$\ge100$ K) atomic 
region comprised of hydrogen, oxygen and ionized carbon
near the surface.  Beyond that, is a cool ($T \sim 50$ K) partially 
dissociated region and still further in the interior is a cooler 
($T \sim 10 - 20$ K) region.  The photo-dissociation regions encompass a wide
range of physical properties.  It appears highly probable that carbon 
recombination lines arise at various depths in the photo-dissociation region.  
From the present results, it appears that ionized carbon can be distributed
in clouds of a range of angular sizes along the line of sight and possess a variety of
physical properties resulting in detectable carbon lines. 

However, the above argument also raises more questions.  If the carbon lines 
arise in the neutral gas,
probably within a low-excitation photo-dissociation region with temperatures ranging from 
80 to 300 K (and possibly 20 K), then why are they not detected 
from many more directions in the Galaxy ?  It appears
reasonable to expect the existence of stronger carbon lines from 
directions which show high \HI
optical depth or high $^{12}$CO emission.  However, this is not always the
case.  For example, inspite of the large \HI optical depth observed 
towards $l=30^{\circ}$ in the Galactic
plane, no carbon line near 328 MHz or 34.5 MHz has been detected.   
Another such interesting direction is towards the extragalactic source 3C~123.  
Although, this direction shows a high \HI optical depth 
($ \tau_{{\mbox{\rm H{\hspace{0.5mm}\scriptsize I }}}} \sim 2.5$), 
no carbon recombination line
near 318 MHz ($n\sim 274$) (\nocite{payne:84}Payne \et, 1984) have been detected down to an 
optical depth limit of $3 \times 10^{-4}$.  \nocite{payne:94} Payne \et (1994)
explain that this non-detection could be due
to the slightly higher temperature of the \HI clouds in this 
direction which render the lines
undetectable.  However, from the modelling in the previous section, 
we find that a range of temperatures ($80 - 300$ K in the
warm models) for the carbon line-forming regions can give rise 
to detectable peak optical depths of $10^{-3}-10^{-4}$ for a 
range of electron densities.  If pressure equilibrium
with \HI gas is assumed, then the models we find have electron densities 
between 0.005 and 0.007 \cm3 ~implying
atomic densities $n_H \sim 50$ \cm3 . 
Larger electron densities ($\sim 0.03$ \cm3 ) that we find for the 
best-fitting models cause the gas to move out of pressure 
equilibrium with \HI. On the other
hand, lower electron densities (\ne $< 0.003$ \cm3 ) require 
extremely long pathlengths through the ionized-carbon gas to 
generate the observed line intensities.  Therefore, if such electron
densities existed in the ionized-carbon gas then also the 
recombination lines would be undetectable.  Hence it appears that 
detectablity of the low-frequency recombination lines is
a sensitive function of both the electron density and temperature.
At this point, it may be useful 
to note that low-frequency observations have detected
carbon lines with line-to-continuum ratios ranging from a few 
times $10^{-4}$ to a few times $10^{-3}$.  It appears certain that 
lines very much stronger than a few times $10^{-3}$ are not common.
However, the lower limit is set by the sensitivity of the 
observations conducted till date. Hence, lines intrinsically weaker 
than few times $10^{-4}$ might possibly exist and would require more
deeper observations to detect them.  It therefore 
appears that with the present sensitivity only 
a subset of the range of values possible for these 
parameters has so far been accessible through
low-frequency recombination lines of carbon.  

The physical properties of the gas in 
the inner Galaxy are remarkably uniform.  Surprisingly, these lines 
which are so widespread in the inner Galaxy ($l < 17^{\circ}$) appear
to be difficult to detect in the outer Galaxy except towards a 
few directions ($l=63^{\circ} \& 75^{\circ}$).  The reason  
possibly lies in the cloud sizes and also in the reduced
background radiation field.  Alternatively, since these
observations have shown that the clouds in the inner Galaxy are 
$\ge 2^{\circ}$, the non-detectability could be due to the 
different physical conditions that may exist in the photo-dissociation regions 
in the inner and the outer Galaxy.  
More sensitive multifrequency observations towards several directions in 
the outer and inner Galaxy are required to ascertain the widespread
existence of low$-\nu$ carbon recombination lines. 

\subsection{The site of formation of the [\CII] 158 $\mu$m line
and the low$-\nu$ carbon recombination lines }
The [CII] $158~\mu$m line is a result of the radiative decay of
the fine-structure transition,
$\rm ^2P_{3/2} \rightarrow {^2P}_{1/2}$ in singly ionized carbon.  The
recombination lines, on the other hand, are a
result of the electronic transitions of an electron
which has most likely dielectronically 
recombined by exciting the same fine structure transition.
The two lines are thus, intimately related and it is natural to 
look for a correlation between the two.  However, 
although dielectronic-like recombination process is one excitation
mechanism for the fine-structure line, it is not the only one.  
Recent advances in infrared observations have
extensively detected the [CII] 158 $\mu$m line from the Galaxy and 
it has been inferred that a wide
range of physical conditions are conducive to the formation of the 
fine-structure line.  $\rm C^{+}$ ions giving rise to the infrared line 
are found to exist in both neutral and ionized regions.
\nocite{shibai:91}Shibai \et (1991) concluded from their balloon-borne experiments
that the diffuse [CII] emission of the Galactic plane comes from the diffuse gas
whereas \nocite{bennett:93}Bennett \& Hinshaw (1993) showed that the [CII] emission 
measured with COBE/FIRAS may originate in the photo-dissociation region.  
\nocite{petuchowski:93}
Petuchowski \& Bennett (1993) and \nocite{heiles:94}Heiles (1994), in separate studies,
analysed the [CII] data and studied its correlation with the 
various possible sites of origin.  Both the studies find that the 
extended low-density warm ionized medium which has a
temperature around $10^4$ K is the main
global contributor to the 158 $\mu$m fine-structure emission line, especially
in the Galactic interior.  \nocite{heiles:94}Heiles (1994) argues that the next 
dominant contributor to the infra-red line is the cold neutral medium and the last 
is the photo-dissociation region.  
On the other hand, low-frequency carbon recombination 
line emission is detected
only from cold neutral gas and probably molecular gas and 
not from the extended low-density warm ionized medium.  
The major contributor to the two species of lines, the \CII $158\mu$m and low-frequency
carbon recombination lines seem to be different
and hence we have not pursued a rigorous treatment of the correlation. 

\section{Summary}
In this paper, we have presented observations of carbon recombination lines
at 34.5 MHz and 325 MHz in the Galactic plane. 
The observations at 34.5 MHz ($n \sim 575$) were conducted using the 
low-frequency dipole array at Gauribidanur
and at 328 MHz ($n \sim 272$) using the Ooty Radio Telescope.  
Carbon lines were detected from nine out of the
32 directions that were observed at 34.5 MHz.  The lines at both the frequencies are weak
with line-to-continuum ratios ranging from a few 
times $10^{-4}$ to $10^{-3}$ and line widths of $15-50$ \kms.  
The similarity of widths observed 
at the two frequencies suggest the
effect of radiation and pressure broadening is negligible and that
the widths are likely due to differential Galactic rotation. 
Observations at 328 MHz, which were made with two angular resolutions,
indicate that the angular size of the line-forming gas is 
$\ge 2^{\circ}$.  Higher resolution VLA observations at 330 MHz 
towards one of the directions, $l=14^{\circ}, b=0^{\circ}$ 
failed to detect any carbon line
emission to a $3\sigma$ level of 75 \mJyb and implies a lower
limit of $10'$ on the angular size of any possible clumps in the line-forming
regions. 

The l-v distribution of the carbon line data 
indicate that the line-forming regions are located between 4 \& 7 kpc
from the Galactic Centre.  Combining the data at
34.5 MHz and 328 MHz with those of \nocite{erickson:95} Erickson \et (1995) at 76 MHz obtained 
using the Parkes telescope, we modeled the line-forming regions.
Upper limits on the electron densities and the
radiation fields were obtained from the observed line widths at 34.5 MHz.  
For different assumed angular sizes of the line-forming regions, 
we found combinations of temperature \Te, density \ne and pathlength $l$
which could satisfactorily explain the observed variation of 
integrated line optical depth with frequency.  The models were found to be  
sensitive to the assumed size of the line-forming region.  In all the models, we 
assumed that the background continuum radiation uniformly fills the beam.  
Models were obtained for
two values of the background radiation field \Tr $=1250$ K and \Tr $=2500$ K.  
For an assumed cloud size $>4^{\circ}$,
we found that the best-fitting parameters are \Te $\sim 20-40$ K, 
\ne $\sim 0.1$ \cm3 ~and $l < 1$ pc.
For an angular size of $4^\circ$, models with \Te ranging from 60 K to 
200 K, \ne between $0.03 - 0.05$ \cm3 and $l$ 
between 4 pc and 90 pc fitted the data.  Finally, if the angular size
of the clouds is $2^{\circ}$ 
then models with \Te ranging from 150 K to 300 K,
\ne $\sim 0.03$ \cm3 and $l$ between 40 pc and 170 pc
could explain the observations. 

The range of possible physical parameters suggest that the
line-forming regions may be associated with photo-dissociation
regions.  Although photo-dissociation regions are known to
be widespread in the Galactic disk, carbon lines are detected
only in about 30\% of the observed positions.  We attribute the paucity
of the detections to (a) limitation of sensitivity, (b) lack of strong
background radiation in some directions and (c) variation in
the physical parameters of photo-dissociation regions.  Further observations of low
frequency carbon recombination lines with higher sensitivity
are likely to yeild many more detections.

\vspace{1cm}

The work was done when NGK was associated with the Raman Research Institute, Bangalore and
the Joint Astronomy Programme, Indian Institute of Science, Bangalore.
We thank A. Santhosh Kumar for the installation and testing of the 8-line receiver and
the operational staff at Gauribidanur for assistance with the long observations.  We thank
Anish Roshi for his generous help with the observations with the Ooty Radio 
Telescope and also for valuable discussions and thank K.S.Dwarakanath for  
help with the VLA observations. The National
Radio Astronomy Observatory is a facility of the National Science Foundation operated under
cooperative agreement by Associated Universities, Inc.

\end{document}